\documentclass[aps,prl,preprint,groupedaddress,nofootinbib]{revtex4}
\usepackage{times}

\usepackage{Packages}
\usepackage{definitions}
\usepackage{comment}

\begin{document}

\preprint{hep-th/0702013}

\title{M-theory compactifications on hyperbolic spaces}


\author{Domenico Orlando}
\affiliation{Universit\`a di Milano-Bicocca  and INFN, Sezione di Milano-Bicocca,  \\
P.zza della Scienza, 3;\\ 
I-20126 Milano, Italy}

\begin{abstract}
  Negatively-curved, maximally symmetric hyperbolic spaces enjoy a
  number of remarkable properties that can be traced back to
  Riemannian geometry, group theory and algebraic geometry. In this note
  we recall some such properties and find $H_n$ as M-theory
  solutions.\\ Based on a talk given at the RTN Workshop 2006, Napoli.
\end{abstract}
\maketitle                   

\newdimen\oldindent
\oldindent=\mathindent

\section{Foreword}
\label{sec:intro}

Maximally symmetric spaces play an important role in supergravity,
because they provide non-trivial, yet manageable solutions. This is
essentially because they allow for a natural description not only in
terms of Riemannian geometry but also in group theory or algebraic
geometry. As a consequence, they enjoy a large number of properties
that make them extremely useful as toy models \emph{e.g.} in
cosmology.

Among these spaces, most of the attention has been devoted to
anti-de~Sitter manifolds and spheres. Here we concentrate our
attention on negatively-curved hyperbolic spaces which can be obtained
in a more non-trivial manner as M-theory solutions, but have already
been considered in
literature~(\cite{Kehagias:2000dg,Gauntlett:2000ng}).


\section{Hyperbolic manifolds}
\label{sec:hyperbolic-manifolds}

Many different definitions can be given for maximally symmetric
spaces. The following is perhaps the most intuitive one. An
$n$-dimensional maximally symmetric space is a pseudo-sphere in $n+1$
dimensions, \emph{i.e.} the locus of the points in $\setR^{n+1}$
satisfying the equation:
\begin{equation}
  \epsilon_0 ( X^0 )^2 + ( X^1 )^2 + \dots + ( X^{n-1} )^2 + \epsilon_n ( X^n )^2 = \epsilon L^2 ,
\end{equation}
where the three $\epsilon$ parameters are signs. In particular
$\epsilon_0$ and $\epsilon_n$ specify the signature of the embedding
$\setR^{n+1}$ space. There are $2^3 - 2 = 6 $ inequivalent choices for
these signs; $(+,+,-)$ gives an empty set and  $(-.-.+)$
results in a manifold with two time directions. The four remaining ones are
summarized in Table~\ref{tab:maximally-symmetric-spaces}.

\begin{table}
  \centering
  \begin{tabular}[c]{c||c|c|c|c|}
    $\left( \epsilon_0, \epsilon_n, \epsilon \right)$ & $---$& $-++$ & $-+-$ & $+++$ \\ \hline \hline
    Signature& Minkowski & Minkowski & Euclidean & Euclidean \\ \hline
    Curvature & $-$ & $+$ & $-$ & $+$ \\ \hline
    Space & $\mathrm{AdS}_n$ & $\mathrm{dS}_n$ & $H_n$ & $S^n$
  \end{tabular}
  \caption{Maximally symmetric spaces}
  \label{tab:maximally-symmetric-spaces}
\end{table}

By construction these spaces have constant curvature. In terms of
Riemannian geometry, the relevant tensors can be written as:
\mathindent=0em
\begin{align}
  R_{abcd} = \frac{\epsilon}{L^2} \left( g_{ad} g_{bc} - g_{ac} g_{bd} \right) \, , &&
  Ric_{ab}  = R\ud{c}{acb} = \frac{\epsilon}{L^2} \left( n - 1 \right) g_{ab} \, ,
  && R = Ric\ud{a}{a} = \frac{\epsilon}{L^2} n \left(n - 1 \right) \, .
\end{align}%
\mathindent=\oldindent%
From now on we put, without loss of generality, $L = 1 $.

In the following, we will mainly concentrate on negatively-curved
Euclidean manifolds, usually called hyperbolic (Poincar\`e,
Lobachevsky) spaces $H_n$. 

Another very useful description can be given in terms of Lie groups. A
maximally symmetric space is identified with the coset $G/T$ (we
quotient with respect to the action of $T$ on the left, $g \sim g t$)
where $T$ is the maximal subgroup in the group $G$. In particular, one
can show that
\begin{equation}
  H_n = \frac{SO(1,n)}{SO(n)} \, .  
\end{equation}
The first obvious consequence is that $SO(1, n)$ is the group of
isometries of $H_n$.

A particularly convenient choice of coordinates, covering the whole manifold, is given by the so-called Poincar\'e coordinates. The line element takes the form
\begin{equation}
  \di s^2 = g_{ab} \di u^a \di u^b = \frac{1 }{(u^1)^2 } \left( (\di u^1)^2 + \dots + (\di u^n)^2 \right)  \, ,
\end{equation}
where $u^i \in \left( 0, + \infty \right)$. In such coordinates it is
evident that such manifolds have infinite volume, \emph{i.e.} the
integral $\left(\int \di u^1 \dots \di u^n \, \sqrt{g} \right)$
diverges.

\bigskip

Starting from $H_n$, it is possible to construct \emph{compact}
manifolds of constant negative curvature by taking the quotient with
respect to the action of a freely acting discrete group $\Gamma$. By
what we said above, $\Gamma$ must be a discrete subgroup of $SO(1,n)$
(from now on we call this a lattice). It is worth emphasizing that,
although $H_n$ and $H_n / \Gamma$ share the same local properties, the
global ones are completely different.\\
Not only that by choosing a lattice in $SO(1, n)$ one obtains a
compact manifold, but it can be shown that $H_n$ is the universal
cover of any closed manifold $M$ of negative constant
curvature. Equivalently, any such $M$ can always be written as $M =
H_n / \Gamma$ for a suitable choice $\Gamma \subset SO(1,n)$.\\
One of the most important results in the study of these manifolds is
\emph{Mostow's rigidity theorem}~\cite{Mostow:1968aa}. It states that
the geometry of a finite volume hyperbolic manifold of dimension
greater than two is determined by its fundamental group\footnote{In
  $d=3$ the constaints imposed by Mostow's theorem can be avoided by
  using Dehn surgery~\cite{Thurston:1978aa}, but the price to pay is
  to use cusped hyperbolic three-manifolds and incomplete metrics. We
  will not deal with this type of varieties in this context.}. More
precisely, we can give two equivalent, geometric and algebraic,
formulations:
\begin{itemize}
\item Given $M$ and $N$, complete, finite-volume hyperbolic
  $n$-manifolds, if there exists an isomorphism $f:\pi_1(M) \to
  \pi_1(N)$, it is induced by a unique isometry from $M$ to $N$.
\item Let $\Gamma_1 $ and $\Gamma_2$ be lattices in $SO(1,n)$ such as
  $H_n / \Gamma_i$ has finite volume. If they are isomorphic then they
  are conjugate.
\end{itemize}
The theorem is not valid in $d=2$ dimensions. This is compatible with
the well known fact that Riemann surfaces of genus $g>1$ (which can
always be represented as quotients $H_2 / \Gamma$, $\Gamma \subset
SO(1,2)$), have a $6 \left( g - 1 \right)$-dimensional moduli space.

In view of the following physical applications, we will now concentrate
on the lower-dimensional $d=2$ and $d=3$ examples.

\paragraph{$d=2$, or, Riemann surfaces.}

Depending on their genus, Riemann surfaces can be endowed with a
metric that can be spherical (genus $g=0$), flat (genus $g=1$) or
hyperbolic ($g\geq2$). This means in particular that any surface of
genus $g > 1$ can be seen as a quotient $H_2/ \Gamma$, where $\Gamma$
is a lattice in $SO(1,2)$.

Any surface of genus $g$ can be described in terms of a polygon. In
particular we define the metric fundamental polygon as the $\left( 4 g
  + 2 \right)$-gon in which the edges are pairwise identified and the
standard fundamental polygon as the $\left(4g \right)$-gon in which
the edges are pairwise identified and all the vertices are identified.
Both the metric and the standard fundamental polygons describe a
surface of genus $g$ as one can easily verify by evaluating the Euler
number that in the first case is $\chi = 1 - \left( 2g + 1 \right) + 2
= 2 - 2g$ (if two vertices are independent) and in the second $\chi =
1 - \left( 2g \right) + 1 = 2 - 2g$. In the usual notation, an $n$-gon
is represented as a string of $n$ letters with exponent $\pm 1$
depending on the orientation of the edge with respect to an arbitrary
positive one. The same letter is used for pairs of identified
edges.
It is possible to show
that the standard fundamental polygon for an oriented Riemann surface
of genus $g$ can always be put in the form
$A_1B_1A_1^{-1}B_1^{-1}\ldots A_g B_g A_g^{-1} B_g^{-1}$.

\begin{figure}
   \begin{center}
  \begin{tabular}{ccc}
    \includegraphics[width=.25\textwidth]{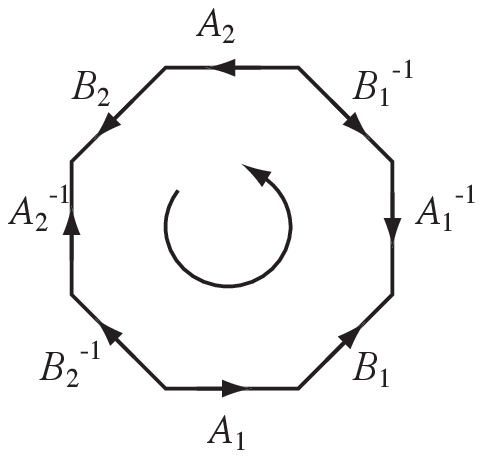} &&     \includegraphics[width=.2\textwidth]{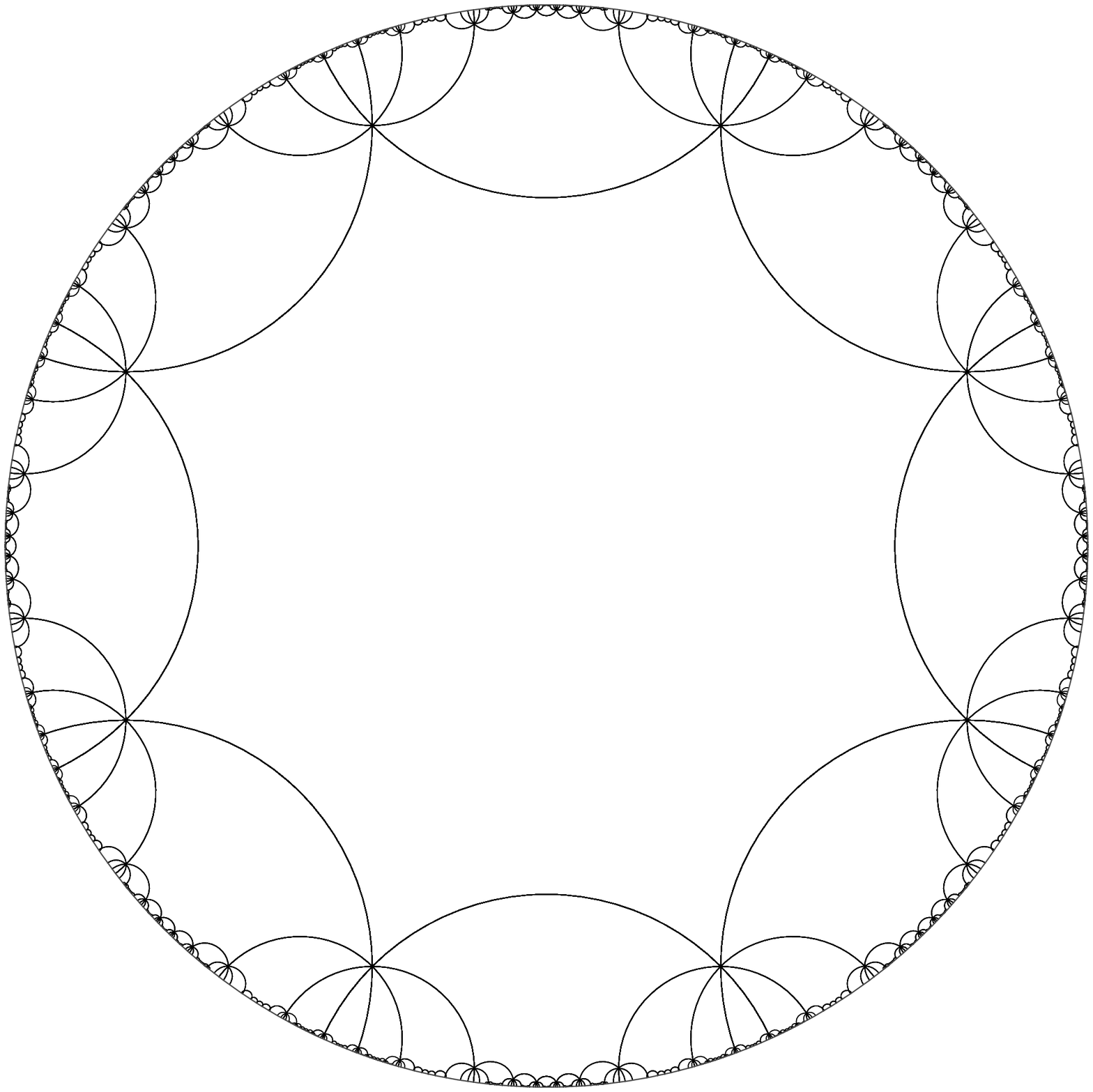} \\
    (a) && (b) 
  \end{tabular}
  \caption{Different representations of a Riemann surface of genus
    $g=2$. (a) Standard fundamental polygon
    $A_1B_1A_1^{-1}B_1^{-1}A_2B_2A_2^{-1}B_2^{-1} $; (b) tessellation
    of the Poincar\'e disc. }
   \end{center}
  \label{fig:Genus-two}
\end{figure}

In an equivalent fashion, the fundamental polygons can be seen as
elementary cells for a tessellation of the hyperbolic plane $H_2$. In
Fig.~\ref{fig:Genus-two} we show one such tessellation, corresponding
to a genus two surface, on the Poincar\'e disc.

\paragraph{Hyperbolic three-manifolds.}

The three-dimensional case is the first example in which Mostow's
theorem holds. In particular, this means that the geometry of a
compact hyperbolic three-manifold is fixed by the choice of the
lattice and hence by its volume. Here we show how this volume can be
computed for a large family of lattices by using algebraic techniques.

The hyperbolic space $H_3$ is the coset $SO(1,3)/SO(3)$. This implies
that its group of isometries is $SL(2, \setC) \sim SO(1,3)$ and in
particular any discrete lattice used to compactify is $\Gamma \subset
SL(2, \setC)$. We want to see how the volume of the manifold can be
evaluated from the lattice. Let $d$ be a square-free integer
(\emph{i.e.} $d$ does not contain any perfect square as
factor). Consider the quadratic field $\setQ (\sqrt{-d}) = \set{a+b
  \sqrt{-d} | a,b \in \setQ}$, \emph{i.e.} the two-dimensional vector
space on $\setQ$ generated by $1$ and $\sqrt{-d}$. Let $\mathcal{O}_d
\subset Q(\sqrt{-d}) $ be the ring of integers in this
field. Explicitly\footnote{When $d = 3 \mod 4$, $\omega $ is an
  algebraic integer since it satisfies the equation $\omega^2 + \omega
  + \nicefrac{\left( d+1 \right)}{4} = 0$.}:
\begin{equation}
  \mathcal{O}_d = \setZ [ \omega], \hspace{1em} \omega =
  \begin{cases}
    \frac{1}{2} \left( -1 + \sqrt{-d} \right) & \text{if $d=3 \mod 4$,} \\
    \sqrt{-d} & \text{otherwise.}
  \end{cases}
\end{equation}
  
The most simple example is $\setQ(\imath)$, the so-called the Gaussian
rationals, \emph{i.e.} the field $\setQ(\imath) = \set{ a + \imath b |
  a,b \in \setQ }$ and $\mathcal{O}_1 $ is the ring of Gaussian
integers $\mathcal{O}_1 = \setZ [\imath]= \set{ a + \imath b | a,b \in
    \setZ}$.

$\mathcal{O}_d$ is a lattice in $\setC$, so we can consider the
lattice $PSL(2, \mathcal{O}_d ) \subset PSL(2, \setC)$ and the
compactifications $H_3 / \Gamma$ where $\Gamma \subset PSL(2, \setC)$.
When one sees $\mathcal{O}_d$ as a lattice in $\setC$, its area is
given by $\vol ( \mathcal{O}_d ) = \sqrt{D}/2 $ where $D$ is the
discriminant:
\begin{equation}
  D (\mathcal{O}_d ) =
  \begin{cases}
    d & \text{if $d=3 \mod 4$,} \\
    4d & \text{otherwise .}
  \end{cases}
\end{equation}
We need now to take into account the difference between considering
the lattice on $\setC$ and on $PSL(2, \setC)$, and this is done by
using the formula~\cite{Sarnak:1983aa}
\begin{equation}
  \vol (H_3 / PSL(2, \mathcal{O}_d))  = \frac{D^{3/2}}{24} \frac{\zeta_{\setQ(\sqrt{-d})}(2)}{\zeta_{\setQ}(2)} \, ,
\end{equation}
where $\zeta_{\setQ} (s) $ is the usual Riemann zeta function and
$\zeta_{\setQ(\sqrt{-d})} (s)$ is the Dedekind zeta function on the
field $\setQ(\sqrt{-d})$.


\section{String and M-theory backgrounds}
\label{sec:string-m-theory}

In this section, we describe the most simple configuration in which
hyperbolic spaces can appear as M-theory backgrounds. Before starting,
it is worth remarking that negatively-curvature Euclidean spaces are
not 'natural' backgrounds. Looking at the structure of the equations
of motion, in fact, one would expect, in the presence of gauge fields
and without arbitrary cosmological constant, to find
negative-curvature Minkowski spaces (anti-de~Sitter) or
positive-curvature Euclidean space (spheres). And, indeed, a no-go
theorem exists for positive-curvature Minkowski spaces
(de~Sitter)~\cite{Maldacena:2000mw}.

Consider the following ansatz for the metric and gauge fields in
eleven dimensional supergravity:
\begin{equation}
  \begin{cases}
    M_{11} = M^{(1)} \times M^{(2)} \times M^{(3)} \, ,\\
    F\form{4} = Q \omega^{(3)} \, ,
  \end{cases}
\end{equation}
where $M^{(i)}$ are maximally symmetric spaces of dimension $d_i =
\dim M^{(i)}$, with the constraint $d_3 = 4$, $M^{(1)}$ has $\left(-,
  +, \dots, + \right)$ signature, $\omega^{(3)}$ is the volume form
for $M^{(3)}$ and $Q$ is a constant. First of all, we remark that the
Chern-Simons contribution to the action $F\form{4} \wedge F\form{4} \wedge
C\form{4}$ is vanishing, and that $\di F_{[4]} = 0 $ since by definition
$\omega^{(3)}$ is covariantly constant. We must then solve only the
Einstein equation:
\begin{equation}
  R_{\mu \nu } - \frac{1}{2} R g_{\mu \nu } - \frac{1}{2} \abs{F\form{4}}^2_{\mu \nu } + \frac{1}{4} \abs{F\form{4}}^2 g_{\mu \nu } = 0 \, ,
\end{equation}
where with the notation $\abs{F\form{p}}^2$ we mean $\abs{F\form{p}}^2_{\mu \nu} = \frac{1}{p!} F_{\mu \mu_2 \ldots \mu_p} F\du{\nu}{ \mu_2 \ldots \mu_p}$ .

Since the $M^{(i)}$'s are symmetric spaces, we can split the Ricci
tensor in blocks and each block will be proportional to the
corresponding metric. To fix the notation we can introduce the
parameters $k_i$ as
\begin{align}
  \left. R_{\mu \nu} \right|_{i} = k_i \left. g_{\mu \nu} \right|_i \, ,&& i = 1,2,3 \, ,
\end{align}
so that the Ricci scalars are given by $R_i = k_i d_i $ and the overall curvature is $R= \sum_i k_i d_i$.

The gauge field is proportional to the volume form, so it is easy to
verify that its square is proportional to the metric:
\begin{equation}
  \abs{F\form{4}}^2_{ab} = Q^2 g_{ab} \, ,
\end{equation}
where $a$ and $b$ only run over the four coordinates of the $M^{(3)}$
manifold.

We hence find that the equations of motion take the form of an
algebraic system for the $k_i$'s:
\begin{align}
\label{eq:k-system}
2k_i -  R = - \left( 1 - 2 \delta_{2,i} \right)
  Q^2 \, ,&& \text{for } i = 1,2,3 \, ,
\end{align}
where $\delta_{2,i}$ is the Kronecker delta.
Before solving the system it is worth to make two remarks that would
remain true also for more general ans\"tze of the same type:
\begin{itemize}
\item Having a minus sign on the \textsc{rhs} implies that negative
  values for $k_i$ are possible.
\item Summing the equations for the $k_i$ one finds that the curvature
  of the Minkowski manifold $M^{(1)}$ is always negative. Or, in other
  words, one cannot obtain de~Sitter compactifications in this
  framework.
\end{itemize}
Solving for $k_i$ we find:
\begin{align}
  k_0 = k_1 = - \frac{Q^2}{3} \, , && k_2 = \frac{2 Q^2}{3} \, ,
\end{align}
and comparing with Table~\ref{tab:maximally-symmetric-spaces}, we find
that these equations describe a space of the type
\begin{equation}
  M_{11} = \mathrm{AdS}_{d} \times H_{7-d} \times S^4 \, .
\end{equation}

\bigskip

The system in Eq.~(\ref{eq:k-system}) only depends on the local
properties of the space and remains valid if instead of $H_{7-d}$ we
consider a $\left(7-d \right)$-dimensional compact hyperbolic manifold
$H_{7-d} / \Gamma$. In this way we find that the solution above also
describes an $\mathrm{AdS}_d$ compactification on a direct product
$H_{7-d} / \Gamma \times S^4 $ which is particularly interesting in
the $d = 4$ and $d=5$ cases.

\bigskip

The next natural step is to study the stability properties of the
solutions we outlined. This goes beyond the scope of the present
note. We will however remark that even if maximally symmetric spaces
have vanishing Weyl tensor and so preserve part of the
supersymmetry, this property will in general be lost in the case of
discrete quotients $H_n / \Gamma$. The existence of Killing spinors
compatible with the boundary conditions has to be checked for each
choice of the lattice $\Gamma$.



\bigskip

\begin{acknowledgements}
  The author thanks C.~Bachas, I.~Bakas, E.~Cremmer, Y.~Dolivet,
  D.~Freedman, C.~Kounnas, M.~Petropoulos, M.~Porrati and S.~Stieberger
  for stimulating scientific discussions. The author is supported in
  part by INFN and MIUR under contract 2005-024045 and by the European
  Community's Human Potential Program MRTN-CT-2004-005104.
\end{acknowledgements}

\bibliography{Biblia}

\begin{thebibliography}{1}

\bibitem{Gauntlett:2000ng}
Jerome~P. Gauntlett, Nakwoo Kim, and Daniel Waldram.
\newblock M-fivebranes wrapped on supersymmetric cycles.
\newblock {\em Phys. Rev.}, D63:126001, 2001.

\bibitem{Kehagias:2000dg}
A.~Kehagias and J.~G. Russo.
\newblock Hyperbolic spaces in string and m-theory.
\newblock {\em JHEP}, 07:027, 2000.

\bibitem{Maldacena:2000mw}
Juan~M. Maldacena and Carlos Nunez.
\newblock Supergravity description of field theories on curved manifolds and a
  no go theorem.
\newblock {\em Int. J. Mod. Phys.}, A16:822--855, 2001.

\bibitem{Mostow:1968aa}
G.D. Mostow.
\newblock Quasi-conformal mappings in n-space and the rigidity of the
  hyperbolic space forms.
\newblock {\em Publ. Math. IHES}, 34:53--104, 1968.

\bibitem{Sarnak:1983aa}
P.~Sarnak.
\newblock The arithmetic and geometry of some hyperbolic three manifolds.
\newblock {\em Acta Mathematica}, 151(1):253--295, 1983.

\bibitem{Thurston:1978aa}
W.~Thurston.
\newblock {\em The Geometry and Topology of Three Manifolds}.
\newblock Princeton Un. Lecture Notes, 1978.

\end{thebibliography}

\end{document}